\documentclass[epjST]{svjour}
\usepackage{graphicx}
\usepackage{graphicx,epstopdf}
\usepackage{graphics}
\usepackage{color}
\begin{document}
\title{Resurgence of oscillation in coupled oscillators under delayed cyclic interaction}
\author{Bidesh K. Bera \thanks{\email{bideshbera18@gmail.com}}, Soumen Majhi \thanks{\email{soumen.majhi91@gmail.com}} \and Dibakar Ghosh \thanks{\email{dibakar@isical.ac.in}}}
\institute{Physics and Applied Mathematics Unit, Indian Statistical Institute, Kolkata-700108, India}

\abstract{
This paper investigates the emergence of amplitude death and revival of oscillations from the suppression states in a system of coupled dynamical units interacting through delayed cyclic mode. In order to resurrect the oscillation from amplitude death state, we introduce asymmetry and feedback parameter in the cyclic coupling forms as a result of which the death region shrinks due to higher asymmetry and lower feedback parameter values for coupled oscillatory systems. Some analytical conditions are derived for amplitude death and revival of oscillations in two coupled limit cycle oscillators and corresponding numerical simulations confirm the obtained theoretical results. We also report that the death state and revival of oscillations from quenched state are possible in the network of identical coupled oscillators. The proposed mechanism has also been examined using chaotic Lorenz oscillator.   } 
\maketitle
\section{Introduction}
\label{intro}
Recently death of oscillation and revival of oscillation from suppressed state of coupled oscillatory systems \cite{phys_rep1,phys_rep2} have become a broad research area due to their potential applications in physical, chemical and biological sciences as well as many other man-made systems. The dynamics of collective behavior of interacting dynamical units have been studied extensively in the last few decades with amplitude death (AD) being one of the important and interesting phenomenon among them. Mainly quenching of oscillations is used as a stabilization control in physical and chemical systems such as laser systems \cite{phys_rep2}, synthetic genetic oscillator \cite{syn_gen_osci}, cellular differentiation \cite{celu_diff} etc. where the oscillation is an unwanted situation. Sufficient time delay in the interaction \cite{Reddy&Sen.PRL.1998} among the oscillators produces AD otherwise oscillators should be non-identical \cite{m1,m2,m3,pla_cyclic}. Besides these, AD is emerged for various coupling forms, for example conjugate coupling \cite{conj}, mean field coupling \cite{mf,mf_srep}, nonlinear coupling \cite{nonlinear}, environmental coupling \cite{env1,env2} and additional repulsive link \cite{repul}. 
\par In various types of natural systems and realistic situations oscillations are omnipresent and in these cases suppression of  oscillations is undesirable. The revival of oscillation is a procedure to restore oscillations from the quenched states while keeping the values fixed for all the parameters of individual oscillators within the network. In this context, feedback is a general mechanism to revive the oscillation from death state \cite{rev2}. The introduction of propagation delay \cite{rev1,rev_epjb} in the interaction term is also capable to resurrect the oscillation from the suppressed states. W. Zou et al. have showed that oscillation can be revived from the death states using {\it feedback parameter} \cite{rev3} in the coupling term. Ghosh et al. examined the restoration of oscillation process from the mean-field \cite{rev4} induced death states of coupled systems. Revival of oscillation scenario is also investigated in a network of coupled oscillators in the presence of direct and indirect interactions \cite{rev_pla}. Very recently this process has been verified experimentally \cite{rev_chaos} using processing delay in interaction term of coupled systems. 
\par In real world systems, time delay interactions are ubiquitous since they are unavoidable due to the finite transmission speeds in many physical, biological and social systems. Emergence of AD due to the time delay interaction has been extensively studied from various aspects. Many types of time delay coupling forms were discovered to be able to produce death state in coupled oscillatory systems such as gradient time-delay coupling \cite{grad}, variable time-delay \cite{vari_delay}, distributed time-delay \cite{dis_delay,dis_delay2}, mean-field repulsive delay \cite{repul_delay}. Global amplitude death occurs in a complex network by increasing the heterogeneity in the coupling delay \cite{hetero_epjb}. Death region is enlarged in complex network for the partial time delay coupling \cite{par_delay} where in the network some oscillators are coupled instantaneously and remaining are time delayed. Also asymmetry delayed coupling \cite{asy_delay} is responsible for the  enhancement of the death region for the lower asymmetry value. For instance, coupling asymmetry occurs naturally in ecosystems where the interaction between the coupled prey and predator systems are always asymmetric \cite{prey_asy}. The effect of the asymmetric coupling leads to several collective phenomena such as enhancement of anomalous phase synchronization \cite{anomalos}, and desynchronization \cite{desyn}.

Again, feedback mechanisms have been used to control diverse dynamics of the governing systems. Also feedback control technique is applied for a powerful treatment of several neuronal diseases such as dystonia, Parkinson's diseases \cite{little}. In the epilepsy research, the processes of suppression of oscillations and revival of oscillations from quenched states have great importance due to their potential applications in electric deep brain simulations for a discharge mitigation \cite{epinsc,epinscmthd,swd,nprneu,gs}. 
On the other hand, beside the suppression of oscillations, it is also important to revive the oscillation from death states since it has wide applications in many physical, biological, environmental and social systems. From the biological point of view, quenched oscillation is fatal in cardiac and respiratory systems \cite{cardiac}, brain waves in neuroscience \cite{neuro} and many other type of physiological processes \cite{physio}, so for the normal functioning of these systems, stable and robust oscillations are very much required. 
\par
\begin{figure}[ht]
	\centerline{
		\includegraphics[scale=0.350]{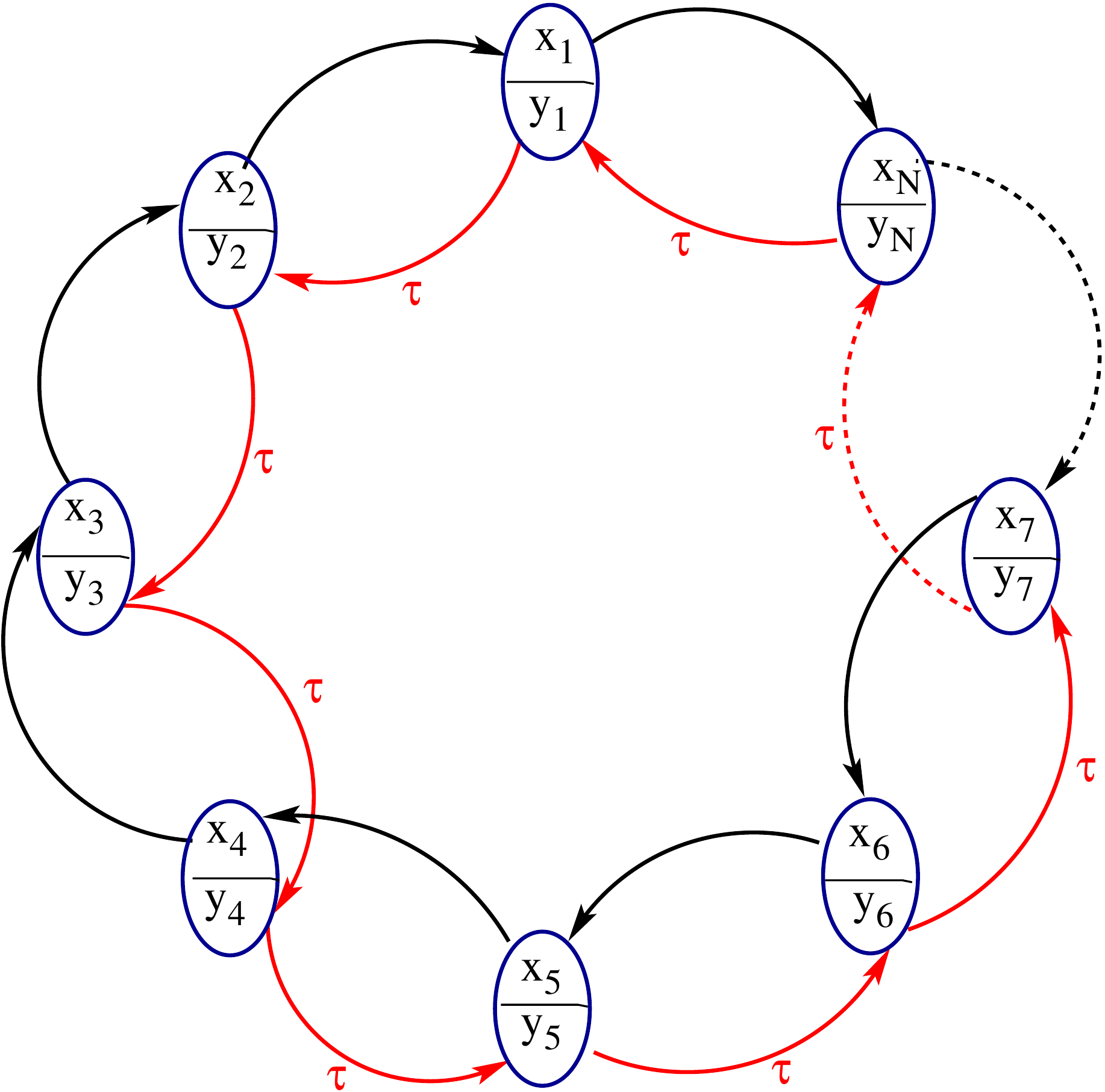}}
	\caption{Schematic diagram of interactions in a network of coupled oscillators in cyclic coupling mode where black and red arrows represent the instantaneous and delayed interactions respectively. The individual oscillators contain two state variables $(x_i, y_i), i=1,2,...,N$ where $N$ is the number of oscillators in the network. Here $\tau$ is the time-delay required to propagate signal from one  oscillator to another through $y$-variable. }
	\label{fig1}
\end{figure}

\begin{figure}[ht]
	\centerline{
		\includegraphics[scale=0.50]{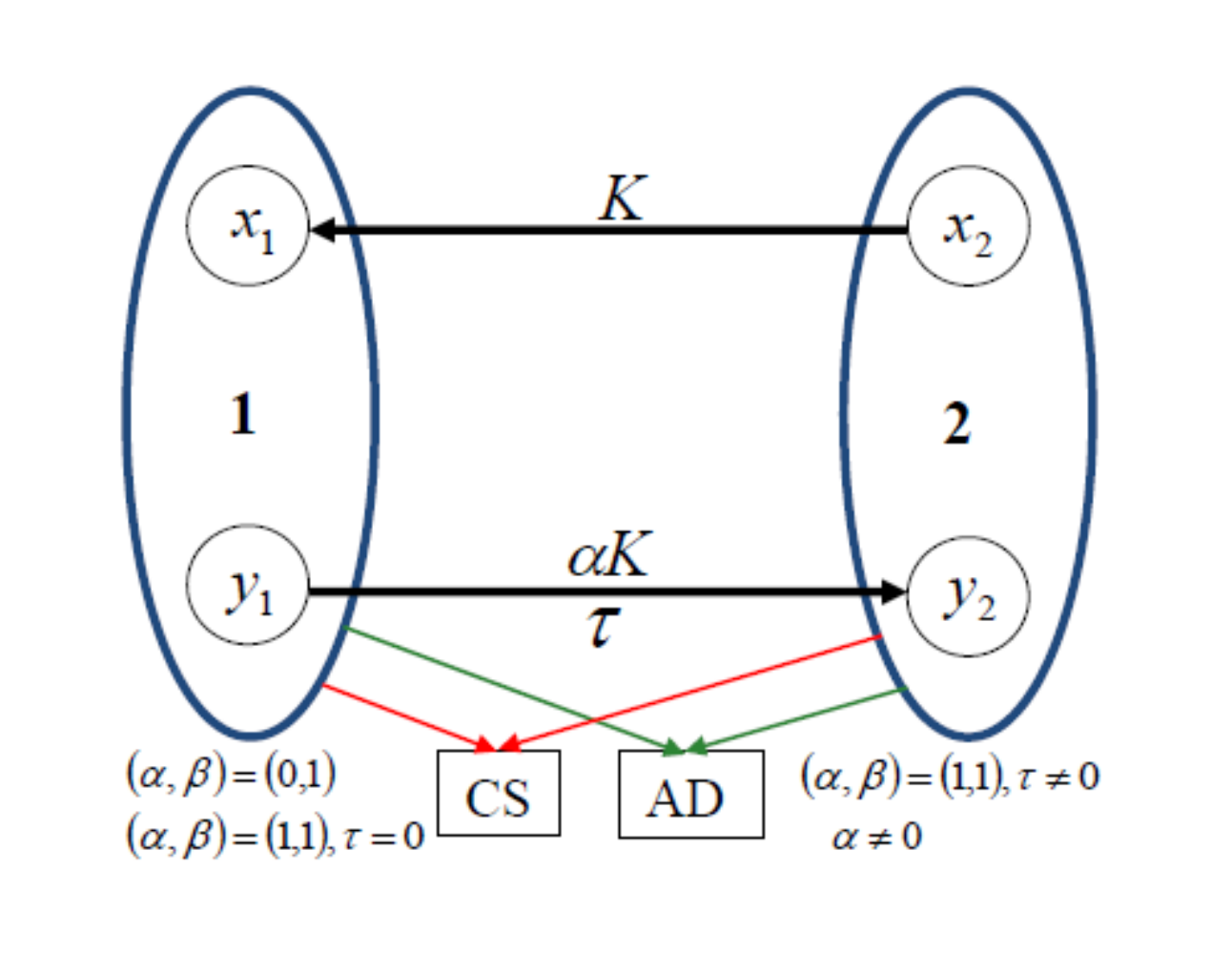}}
	\caption{Cyclic coupling configuration between two coupled systems and various type of collective dynamics that occur depending on different controlling parameters.}
	\label{fig2}
\end{figure}

 The effect of time delay in the coupling function plays a crucial role to produce AD \cite{Reddy&Sen.PRL.1998,physica_D} and its evidence is proved in laboratory experiment \cite{Reddy&Sen.PRL.2000}. Most of the previous works are confined to homogeneous time delay or distributed time delay \cite{dis_delay,dis_delay2} coupling for the occurrence of AD phenomena in globally and locally coupled oscillators by considering the network of identical as well as nonidentical oscillators. Here we use asymmetric cyclic delayed coupling form which is a \emph{pull-push} type interaction between the oscillators. Now delayed cyclic coupling is defined as a mutual interaction between two oscillators where first oscillator sends a signal through  one pair of state variables to second oscillator and receives feedback through another pair of state variables after some time. Where the first two state variables are instantaneously coupled, the latter pair of state variables are delay coupled. Here we induce  asymmetry in delayed interaction which means the instantaneous coupling strength differs from the delayed coupling strength by multiplication of a constant. We explore the effect of feedback parameter in coupled systems. The schematic coupling strategy in a driver and response systems is shown in Fig.~\ref{fig1}, where black and red arrows respectively represent the instantaneous and delayed interactions in the network of coupled oscillators with cyclic configuration. The proposed cyclic coupling configuration between two coupled systems and various types of collective dynamics occurs depending on different control parameters has been explicitly shown in Fig~\ref{fig2}. In the absence of coupling time delay, the transition from amplitude death to oscillation death \cite{pla_cyclic} was observed in coupled systems where negative parameter mismatch played the crucial role.
 Cyclic type of topology is detected in polymer network \cite{cyclic_polymer} and also the cyclic defects mechanism is used for suppressing the spread of information in many real networks from routing loops in computer networks and acquaintance clusters in social networks to many forms of network in science and engineering. The most simple example of the cyclic interaction is the children's rock-scissors-paper game where rock crushes scissors, scissors cut paper, and paper covers rock. This type of interaction is usually seen in neuronal systems where one neuron sends a signal through a pair of dendrites and receives the feedback after some time via another pair of dendrites \cite{cyclic_book}. The mating strategy of side-blotched lizards \cite{lizard}, overgrowths by marine sessile organisms \cite{marin_selism}, and competitions among various strains of bacteriocin producing bacteria \cite{bactaria} are the cyclic type of interaction. The competition in microbial populations are the cyclic dominance \cite{cyclic dominance1,cyclic dominance2} type of interaction and also these types of interactions occur spontaneously in various types of predator-prey interactions and different types of evolutionary games where the competing strategies are three or more. Recently, C. W. Feldager et al. \cite{cyclic_pre} studied the cyclically competing species model on a ring where the network is characterized by Lotka-Volterra equation and they showed that the competing species extinct deterministically with a global mixing at finite rate.
\par In this paper, we introduce two control parameters, namely  feedback and asymmetry parameter to revive oscillation from the quenched state where AD emerged due to sufficient amount of delay in cyclic interaction in the coupled oscillators. For coupled oscillators, in the asymmetric parameter case death region decreases in the parameter space of coupling strength and time delay due to higher values of asymmetric parameter with unit feedback value and opposite scenario is observed in the feedback parameter case. We derived the critical stability curves which bound the death region in the phase space of coupling strength and time delay for coupled systems. The theoretical results have been done on coupled Landau-Stuart oscillators and obtained numerical results support our theoretical results. A numerical justification of feedback case for coupled Lorenz systems has also been given.
\section{ COUPLED LANDAU-STUART OSCILLATORS }

The prescribed coupling scheme (as in Fig. 1) is illustrated using Landau-Stuart (LS) limit cycle oscillators. The mathematical form is given by: \\\\$ \dot x_i=(1-{x_i}^2-{y_i}^2)x_i-\omega y_i+K(x_{i+1}-\beta x_i) $\\$
\dot y_i=(1-{x_i}^2-{y_i}^2)y_i+\omega x_i+K \alpha (y_{i-1}(t-\tau)-\beta y_i),$$$ \eqno{(1)}$$
where $i= 1,2,...,N(\geq3)$, $N$ being the number of oscillators in the network with  $ x_{N+1}(t)=x_1(t) $ and $ y_0(t-\tau)=y_N(t-\tau) $, $\alpha $ is the asymmetry in the coupling strength and $\beta$ is  the feedback parameter. In the absence of coupling, an isolated oscillator has an unstable focus at origin and stable limit cycle at $x_i^2+y_i^2=1$. Here $\omega$ is the intrinsic frequency of the individual oscillators and $K$, $\tau$ are the coupling strength and time delay respectively. If $\alpha =0$ and $\beta=1$, then the oscillators of network are unidirectional diffusively coupled and for certain coupling strengths they get completely synchronized (CS) but never produce AD with identical frequencies. In the case of $\alpha =1$ and $\beta=1$, oscillators are cyclically coupled with time delay $\tau$ that can produce AD for suitable value of coupling strength $K$. In the present case, we fix the value of feedback parameter $\beta$ as unity. When we introduce the asymmetry $\alpha$ in delayed cyclic mode then AD  occurs and depending on the asymmetry value, the size of death island is found to vary and oscillations are revived. Now again if we tune the value of $\beta$ (keeping $\alpha$ fixed as unity) then the size of death region again changes and restoration of oscillation is observed from a suppressed state.  Described cyclic interaction between two ($N=2$) LS oscillators is given by the following dynamical equations
	\\ $ \dot x_1=(1-{x_1}^2-{y_1}^2)x_1-\omega y_1+K(x_{2}-\beta x_1) $\\
	$\dot y_1=(1-{x_1}^2-{y_1}^2)y_1+\omega x_1$ \\
	 $ \dot x_2=(1-{x_2}^2-{y_2}^2)x_2-\omega y_2 $\\
	 $\dot y_2=(1-{x_2}^2-{y_2}^2)y_2+\omega x_2+K\alpha(y_1(t-\tau)-\beta y_2).$$$\eqno{(1a)}$$ We linearize the above coupled systems around the equilibrium point origin and the corresponding characteristic equation is
$$\{(1-\lambda)(1-K\alpha\beta-\lambda)+\omega^2\}\{(1-\lambda)(1-K\beta-\lambda)+\omega^2\}+K^2\omega^2\alpha e^{-\lambda\tau}=0.\eqno{(1b)}$$

In the presence of both $\alpha$ and $\beta$, it is very difficult to obtain critical stability curve analytically.  In the next section, we derive analytically the critical stability curves for certain values of $\alpha$ and $\beta$.  

\subsection{ Asymmetrically coupled Landau-Stuart oscillators}
For simplicity we consider two coupled LS oscillators with delayed cyclic coupling and find the critical stability curve analytically for $N=2$. First we discuss the symmetry case and without feedback parameter, i.e. for $\alpha=\beta=1$. The system of two cyclically delayed coupled LS has a trivial fixed point $(0,0,0,0)$ and we analytically derive the critical stability curves for $\alpha=1$ and $\beta=1$ which bound the death island. After linearizing the coupled systems at origin, the characteristic equation (1b) becomes
\\$$[(1-\lambda )(1-\lambda-K)+\omega^{2}]^{2}+K^{2}\omega ^{2}e^{-\lambda \tau }=0,\eqno{(2)}$$\\ where $\lambda $ is, in general, a complex eigenvalue. From the above equation we get two equations for $\lambda $ \\$$(1-\lambda )(1-\lambda-K)+\omega^{2}+iK\omega e^-\frac{{\lambda \tau }}{2}=0 \eqno{(2a)}$$ and \\$$(1-\lambda )(1-\lambda-K)+\omega^{2}-iK\omega e^-\frac{{\lambda \tau }}{2}=0. \eqno{(2b)}$$\\The stability depends on the nature of the real part of the complex root $\lambda $. For AD the real parts of $\lambda$ should be negative. Thus stability changes when real part crosses the imaginary axis as the parameters vary. Thus for Hopf bifurcation condition we put $\lambda =i\lambda_{1}$. After simplification we get the following critical curves which bound the death region,\\$$\tau_{1}=\frac{2}{p_{2}}\cos^{-1}(\frac{bp_{2}}{c})\eqno{(3a)}$$ and $$\tau_{2}=\frac{2}{p_{1}}(\pi-\cos^{-1}(\frac{bp_{1}}{c}))\eqno{(3b)}$$ \\where $p_{1}=\frac{1}{\sqrt{2}}\sqrt{(2a-b^{2})+\sqrt{b^{4}-4ab^{2}+4c^{2}}}$ and\\
\begin{figure}[ht]
	\centerline{
		\includegraphics[scale=0.50]{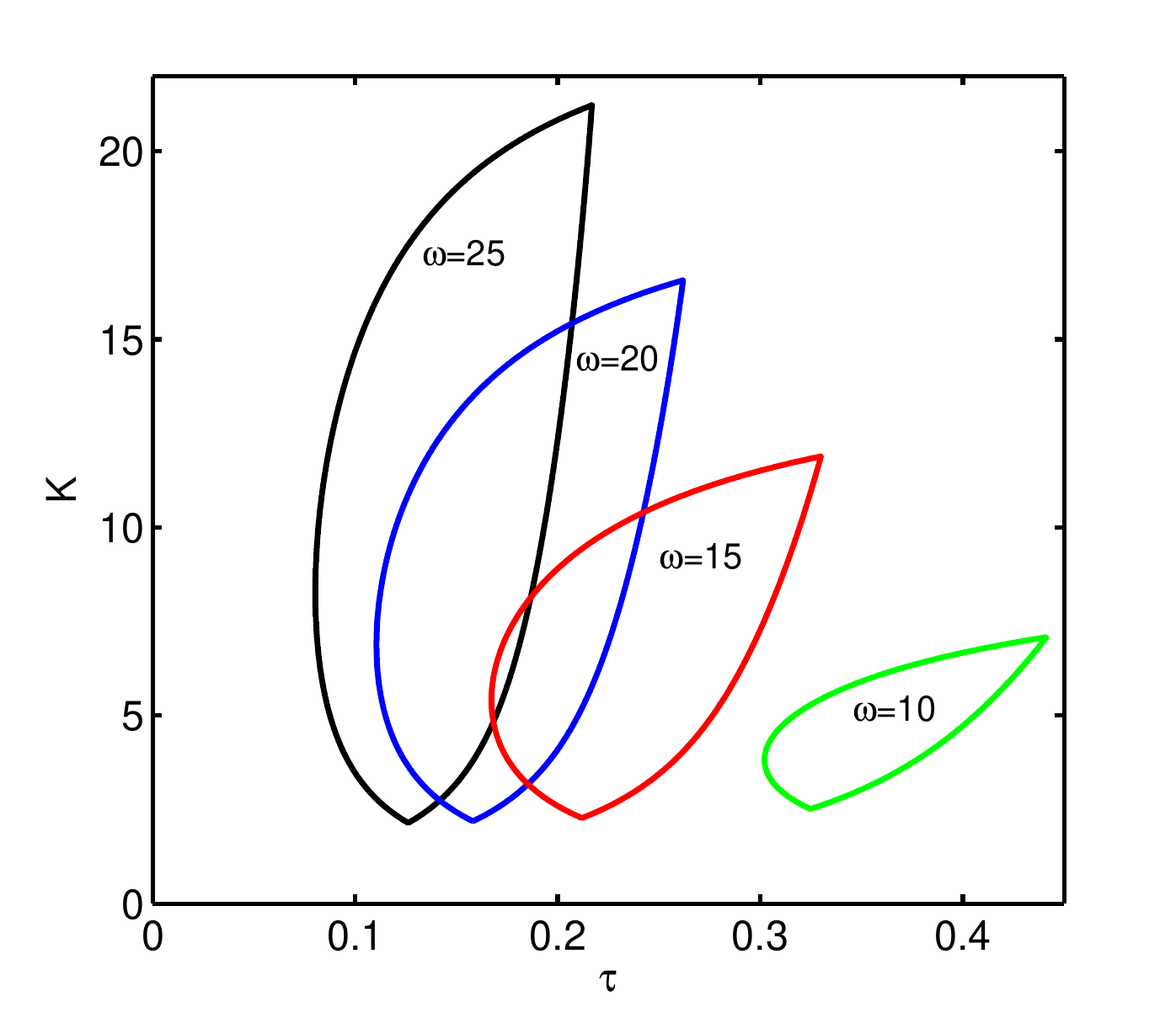}}
	\caption{(Color online) Amplitude death islands for two coupled LS oscillators with unit feedback and asymmetry parameters for different values of $\omega=10, 15, 20, 25$ in the space of coupling strength $K$ and time delay $\tau$.}
	\label{fig3}
\end{figure}

 $p_{2}=\frac{1}{\sqrt{2}}\sqrt{(2a-b^{2})-\sqrt{b^{4}-4ab^{2}+4c^{2}}}$, $a=1-K+\omega^{2}$, $b=K-2$ and $c=K\omega$.\\In Fig.~\ref{fig3}, we plot the AD islands bounded by the critical stability curves ($3a$, $3b$) in the $(\tau , K)$ plane for different values of $\omega$, which are obtained analytically. The death islands bounded by black, blue, red, and green curves correspond to $\omega=25$, $\omega=20$, $\omega=15$, and $\omega=10$ respectively. Now in order to revive oscillation from death states, one should tune the parameters like $\alpha$ or $\beta$ keeping the intrinsic frequency $\omega$ at fixed value. Henceforth, we fix $\omega=15$ throughout our work.    
But when we induce asymmetry in the interaction, it is difficult to obtain the critical stability curve analytically, so we verify it numerically. To integrate the coupled delayed differential equations, we use the modified Heun method \cite{huen} with integration step length $\Delta t=0.01$. The initial conditions for each oscillator in the network (1) are chosen as completely random which are uniformly distributed in the range [$-1,1$] for $-\tau\le t \le 0$.
\begin{figure}[ht]
	\centerline{
		\includegraphics[scale=0.650]{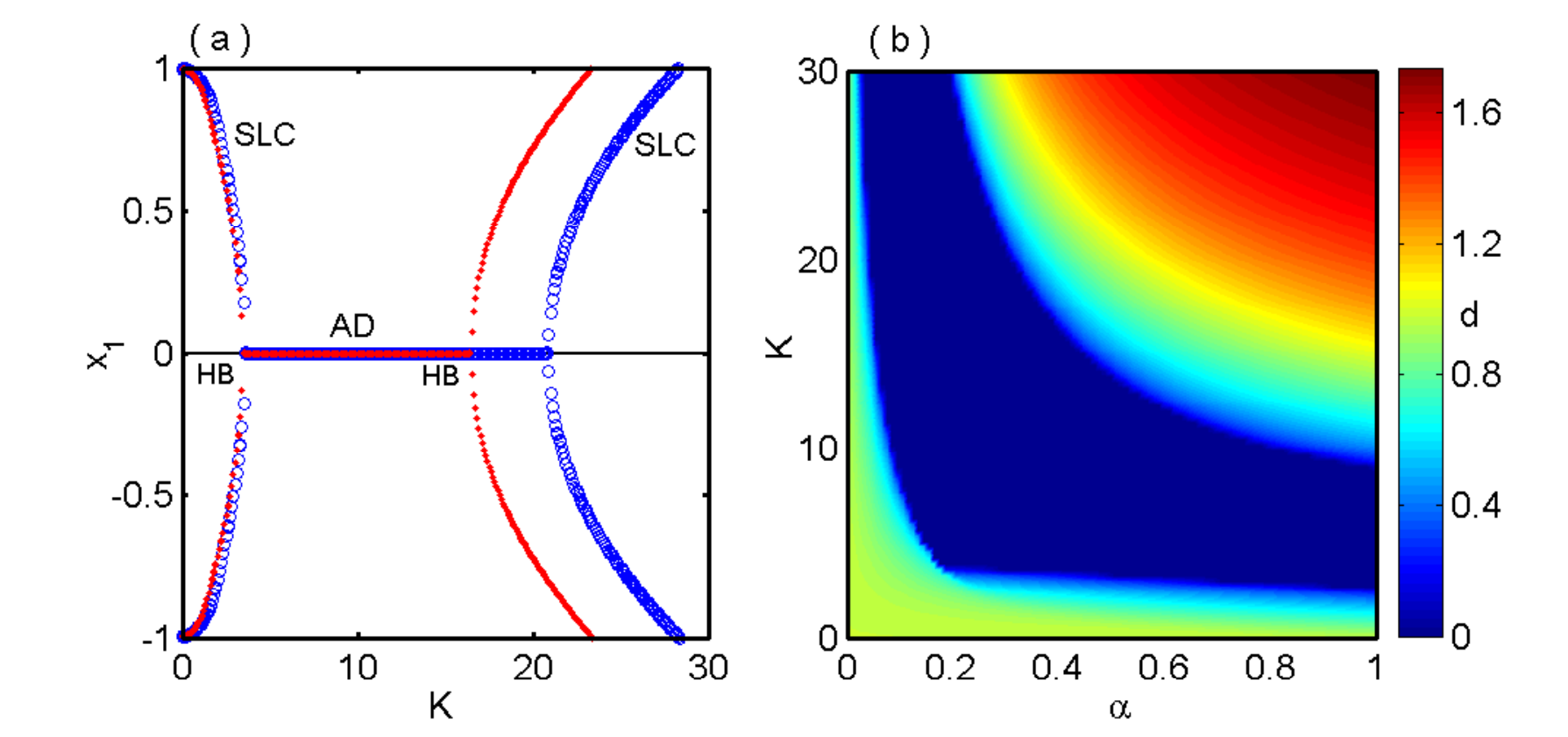}}
	\caption{(Color online) (a) The bifurcation diagram with respect to $K$ showing the transition from SLC to  AD and the restoration of oscillation from the suppressed state for fixed $\beta=1,\tau=0.2$ and different asymmetry values $\alpha=0.3$ (blue), $\alpha=0.4$ (red). (b) Transition from the oscillation state to death state and vice versa in the parameter plane of interaction strength $K$ and asymmetry parameter $\alpha$ with $\tau=0.20$, $\beta=1$ of coupled LS oscillators is plotted, where average amplitude parameter $d$ is used as a color bar. Death region is represented by the deep blue (black) region and the rest of the regions correspond to the oscillatory states.}
	\label{fig4}
\end{figure}
\par To characterize the AD state, we introduce an order parameter which measures the average amplitude of the coupled oscillators and is defined as $d=\frac{d_1+d_2}{2}$, where $d_i=\langle x_{i,max}\rangle_t-\langle x_{i,min}\rangle_t $ for $i=1,2$ and $\langle ... \rangle_t$ represents sufficient long time average.  To calculate the order parameter and death region, the time interval is taken over $4\times10^5$ time units after an initial transient of $1\times10^5$ units. Clearly, the defined quantity $d=0$ refers to the AD state whereas nonzero values correspond to the oscillatory states. The transition from stable limit cycle (SLC) to AD and restoration of oscillation from the quenched state are shown in Fig.~\ref{fig4}(a) by drawing the bifurcation diagram against the coupling strength $K$ with fixed $\beta=1.0,\tau=0.2$ where the blue and red colors correspond to  $\alpha=0.3$ and $\alpha=0.4$. The transition from SLC to AD appears through Hopf bifurcation (HB) at the same coupling values $K=3.5$ for different values of $\alpha$ while the restoration of oscillation from AD state occurs via HB at lower value of $K=16.2$ (red) compared to $K=20.8$ (blue) that happens due to the increasing values of asymmetry parameter $\alpha$. In Fig.~\ref{fig4}(b) we show the variation of the death island in the $(\alpha, K)$ phase space with $\beta=1$. In this figure blue color corresponding to $d=0$ represents the AD state while all the other colors associated with non-zero $d$ signify the oscillatory states. From Fig.~\ref{fig4}(b), one can see that for small values of $\alpha$ (say $\alpha=0.1$), AD occurs at $K=9.3$ and persists for a long range of $K$. But as $\alpha$ increases to $\alpha=0.4$, the death region persists for $3.5<K<16.2$ and oscillation gets revived from the suppressed state in that region. For further larger values of a, for example $\alpha=0.8$, AD exists in smaller range of $K$ ($3.0<K<10$). In this way, asymmetry parameter $\alpha$ helps to shrink AD region in $\alpha-K$ space,  thus oscillations are revitalized.

\par
\begin{figure}[ht]
 	\centerline{
 		\includegraphics[scale=0.50]{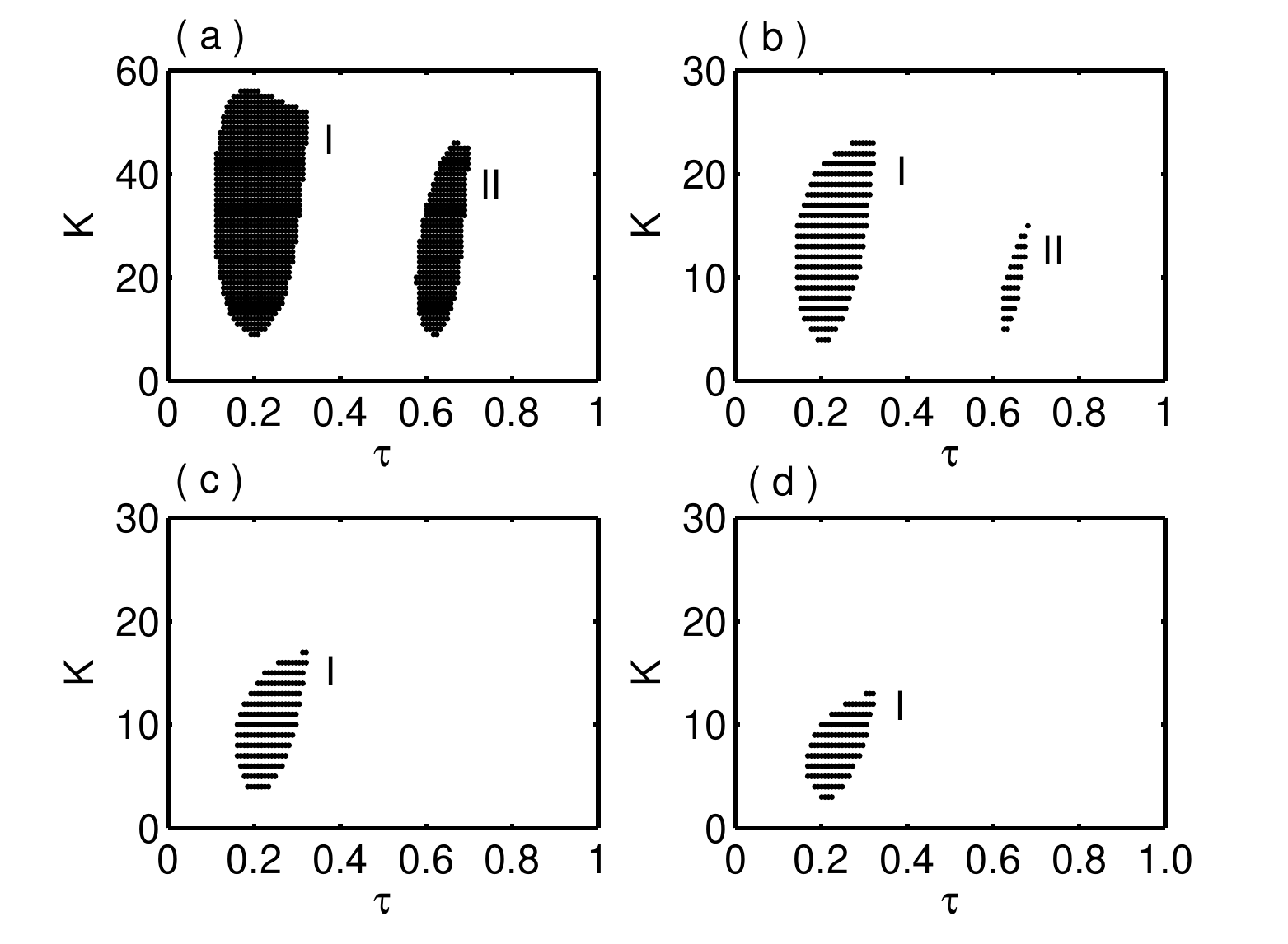}}
 	\caption{ The death islands in the $(K, \tau)$ parameter space for coupled oscillators with fixed frequency $\omega=15$, feedback $\beta=1$ and different values of asymmetry parameter $\alpha$: (a) $\alpha=0.1$, (b) $\alpha=0.3$, (c) $\alpha=0.5$, and (d) $\alpha=0.8$. } 
 	\label{fig5}
 \end{figure}

 In Fig.~\ref{fig5}(a), the death regions (indicated as I and II) for $\beta=1$ and $\alpha=0.1$ is plotted in the $\tau-K$ space. Now as we increase $\alpha$ slightly at $\alpha=0.3$, the size of both regions I and II decrease in the parameter space (Fig.~\ref{fig5}(b)) and thus oscillation is restored from the death state. Further increment in $\alpha$ to its higher value  of $\alpha=0.5$ leads to complete disappearance of region II and shrinkage of region I as depicted in Fig.~\ref{fig5}(c), and thus oscillation is again revived for certain values of the parameters $K$ and $\tau$. When $\alpha=0.8$, the further shrunken death region is shown in Fig.~\ref{fig5}(d) and the oscillation is resurrected in a wide range of the parameter space. This is how increasing the value of the asymmetry parameter $\alpha $ (with $\beta=1$) the stable death region decreases with fixed frequency $\omega=15$,  hence oscillation is resurrected from the AD state. 

\par
\begin{figure}[ht]
	\centerline{
		\includegraphics[scale=0.50]{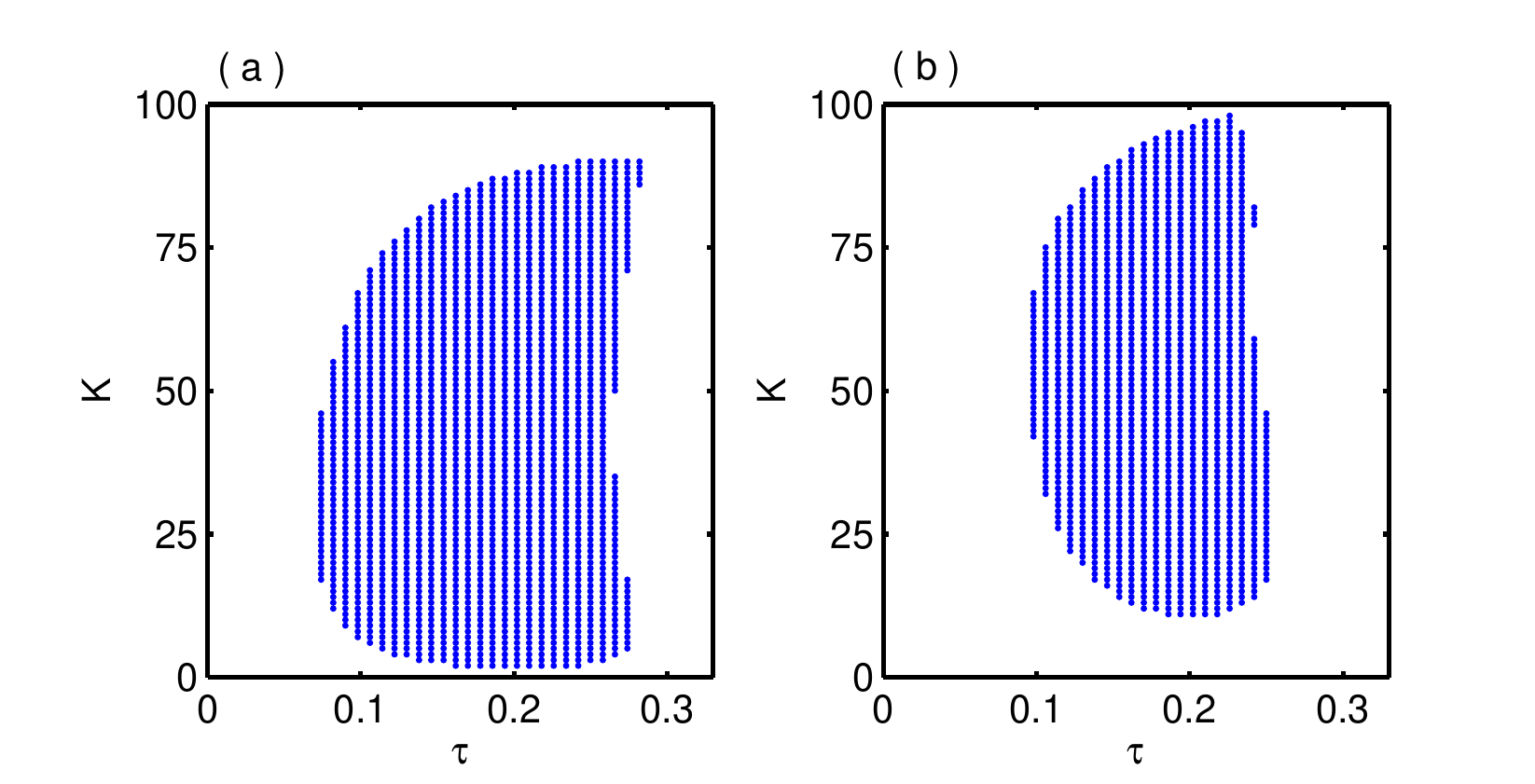}}
	\caption{(Color online) The amplitude death region in the space of interaction strength $K$ and coupling delay $\tau $ in the network of asymmetrically coupled LS oscillators. Figs (a) and (b) represents the death regions for different values of $\alpha= 0.1$ and $0.95$ respectively for fixed value of $\omega=15$, $\beta=1$ and network size $N=100$.}
	\label{fig6}
\end{figure}

 Next we consider network of oscillators interacting through asymmetry delayed cyclic form with $\beta=1$. The variation of the death island in the  network of coupled systems are shown in Fig.~\ref{fig6}, where we take $N=100$ oscillators having identical frequency $\omega=15$ and changing value of asymmetry parameter $\alpha$ in $(K,\tau)$ parameter space. Figs.~\ref{fig6}(a,b) represents the AD islands by simultaneously varying the value of $\tau$ and $K$ for $\alpha=0.1, 0.95$ respectively. From these figures it is clear that the death region decreases due to the increasing values of the asymmetry parameter $\alpha$ in the coupling strength and oscillations are reanimated for certain values of $K$ and $\tau$.

\subsection{Coupled Landau-Stuart oscillators with feedback}
\begin{figure}[ht]
	\centerline{
		\includegraphics[scale=0.550]{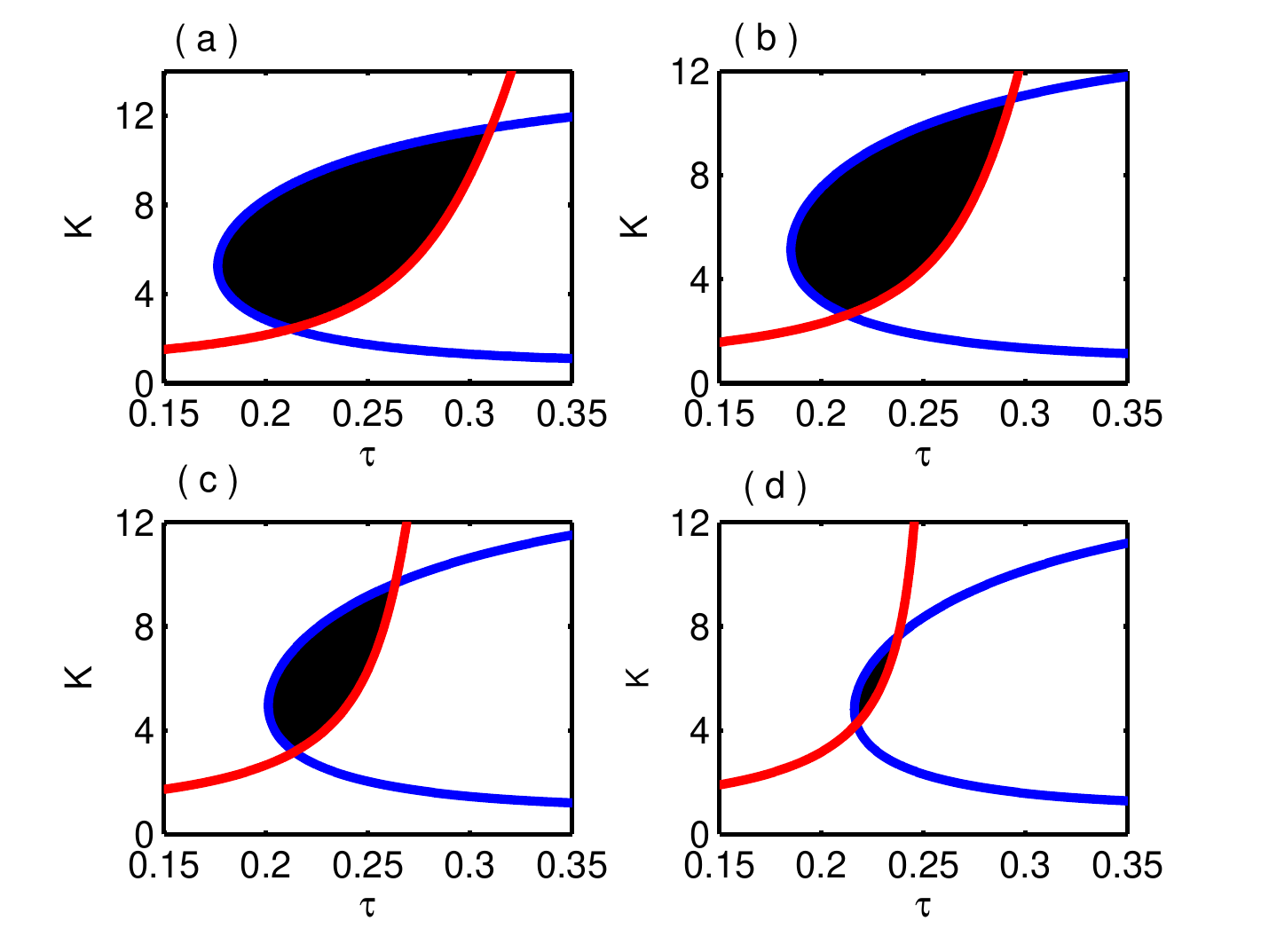}}
	\caption{(Color online) The critical stability curve of two coupled LS oscillators bounding the amplitude death island in the space of coupling strength $k$ and delay time $\tau$ for different values of a feedback parameter $\beta$: (a) $\beta=0.95$, (b) $\beta=0.9$, (c) $\beta=0.8$ and (d) $0.7$. Red and blue color solid curves represent the analytical curves (4a) and (4b) respectively. Black region bounded by analytical curves (4a) and (4b) are obtained using numerical simulation. }
	\label{fig7}
\end{figure}

Next we consider the two cyclic coupled oscillators where a parameter is induced as a feedback in the coupling form and it plays a crucial role in  controlling the AD and oscillation states. From equation (1a) for unit asymmetry value ($\alpha=1$) and in the presence of feedback parameter ($0<\beta<1$) for two coupled oscillators, there exists a fixed point at origin and the corresponding characteristic equation (1b) about the trivial equilibrium point $(0,0,0,0)$ is\\ $$((1-\lambda)(1-\beta K -\lambda)+\omega^2)^2+K^2\omega^2e^{-\lambda\tau}=0\eqno{(4)}$$\\ and the corresponding stability curve for the AD is\\$$\tau_1=\frac{2}{\beta_2}\cos^{-1}(\frac{b\beta_2}{c})\eqno{(4a)}$$\\
$$\tau_2=\frac{2}{\beta_1}(\pi-\cos^{-1}(\frac{b\beta_1}{c}))\eqno{(4b)}$$\\

where $\beta_1=\frac{1}{\sqrt{2}} \sqrt{(2a-b^2)+\sqrt{b^4-4ab^2+4c^2}}$\\
and $\beta_2=\frac{1}{\sqrt{2}} \sqrt{(2a-b^2)-\sqrt{b^4-4ab^2+4c^2}}$\\
with $a=1-K \beta+\omega^2, b=K\beta-2, c=K\omega. $\\

\begin{figure}[ht]
	\centerline{
		\includegraphics[scale=0.50]{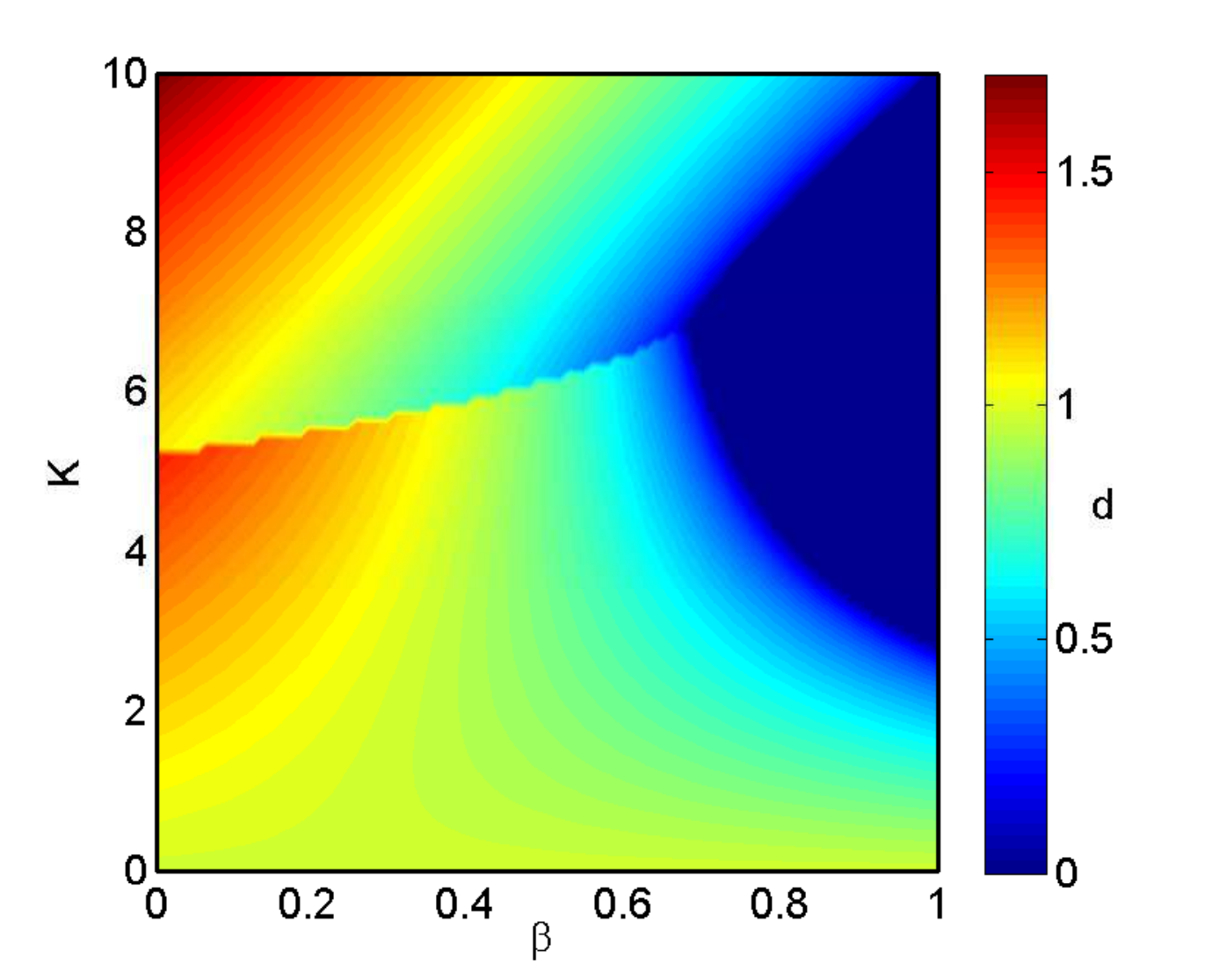}}
	\caption{(Color online) In the $\beta-K$ space the transition from oscillation state to death state is shown for fixed $\tau=0.23$ and $\alpha=1$ parameters in coupled LS oscillators, where average amplitude parameter $d$ is used as a color bar. Death region is represented by the deep blue (black) region and the rest of the regions correspond to the oscillatory states.}
	\label{fig8}
\end{figure}

In Fig.~\ref{fig7} the AD islands bounded by the critical stability curves in the $(\tau , K)$ plane for different values of $\beta$ have been plotted with fixed frequency $\omega=15$. The critical stability curves are derived analytically as above and are drawn by red and blue curves. The regions bounded by those curves in black color are numerically obtained that clearly fit with the analytical results. The death islands for $\beta=0.95, 0.9, 0.8$ and $0.7$ are shown in Figs.~\ref{fig7}(a-d) respectively. Thus the death regions can be seen to shrink drastically in the parameter space $(\tau , K)$ for even a feeble decrement in the value of $\beta$ from unity and this is how oscillations are revived in the corresponding region. Further decrement in the value of $\beta$ leads to complete annihilation of the death state in the parameter space.
In Fig.~\ref{fig8} the variation of the death island is shown in the $(\beta, K)$ space for $\alpha=1$. From this figure, it is noticed that when $\beta=1$, amplitude death exists for the range of coupling strength $2.9<K<10$ but as $\beta$ starts decreasing, even a tiny deviation from unity is making the range of coupling strength (for which death occurs) much smaller than the previous one. Therefore the oscillation is getting revived for smaller values of $\beta$ and $\beta=0.6$ makes both the oscillators to oscillate again for any value of coupling strengths.  If we introduce the coupling delay $\tau$ in the equation of $x$-variable instead of the equation of $y$-variable, then the qualitative and quantitative parameter space for AD and oscillatory states in $\tau-k$ and $\beta-k$ planes remain exactly the same as there is no time-scale separation between $x$ and $y$.    

\par
\begin{figure}[ht]
	\centerline{
		\includegraphics[scale=0.650]{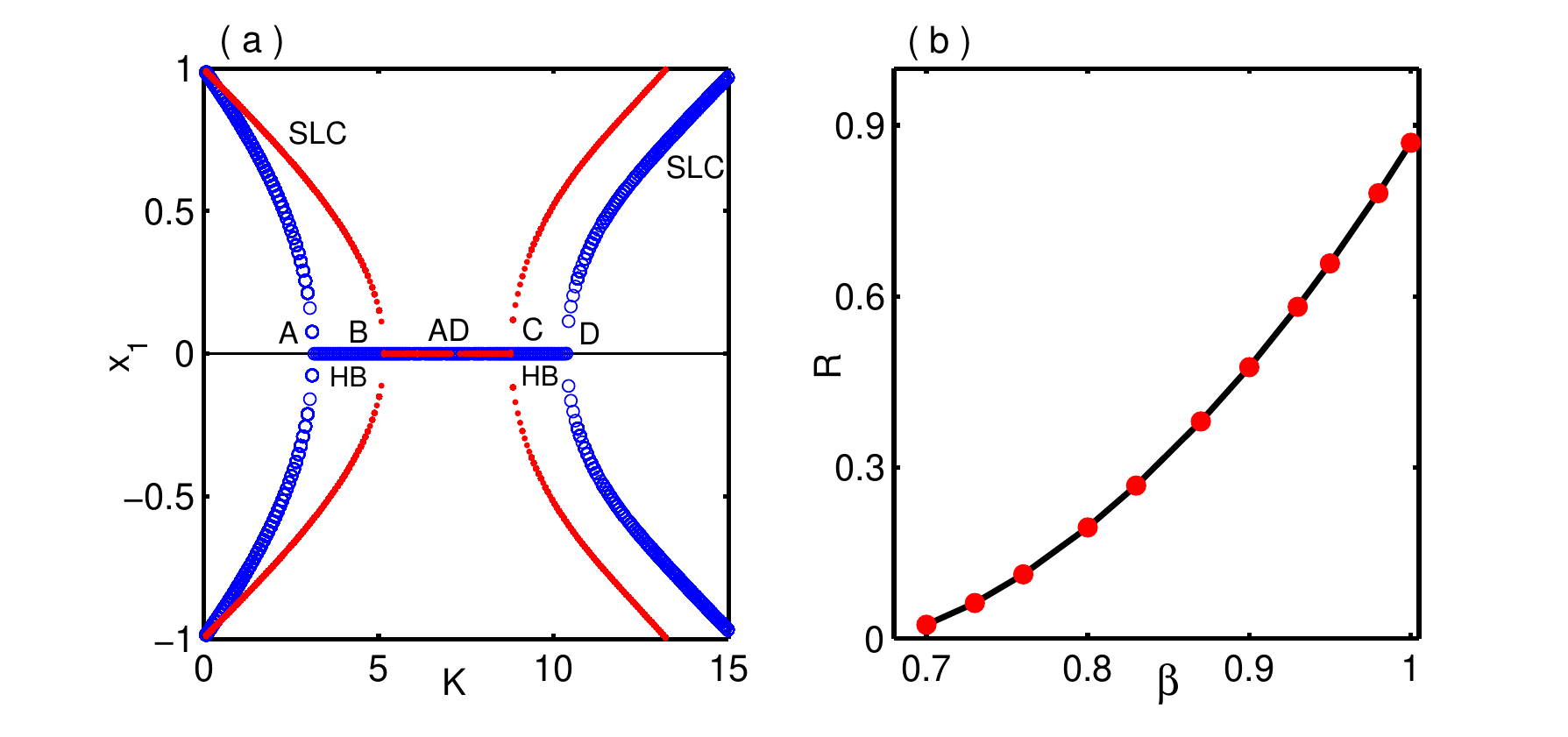}}
	\caption{(Color online) (a) Bifurcation diagram of two coupled LS oscillators where blue curve represents $\beta=1$ and red curve represents $\beta=0.8$ where  $\tau=0.24, \omega=15$ and $\alpha=1$. (b) For fixed frequency $\omega=15$ and $\alpha=1$, the change of area $R$ of the death region with respect to feedback parameter $\beta$.}
	\label{fig9}
\end{figure}

 To give a detailed verification of the scenario of resurrection of oscillation for the feedback case discussed above, we draw bifurcation diagrams taking $K$ as the bifurcation parameter in Fig.~\ref{fig9}(a). The blue open circle curve represents stable limit cycle (SLC) and blue line the stable steady state which refers to the AD state, occurring and disappearing at `A' and `D' respectively through supercritical Hopf bifurcations (HB) for $\beta=1$. In this case AD exists for the range $3.17<K<10.35$ of the coupling strength $K$. However a slight decrement in $\beta$ value to $\beta=0.8$ gives the red dotted curve as the SLC and the red line as the stable steady state corresponding to AD (again occurring at `B' and disappearing at `C' through supercritical HB ). This time AD persists for $5.21<K<8.71$ meaning shorter $K$ range produces AD leading to the revival of oscillations. To quantify the variation of the AD region with respect to the feedback parameter $\beta$, we compute a quantity $R$ which denotes the area of the AD island in the $K-\tau$ space bounded by the critical stability curves. We calculate $R$ by taking different values of $\beta$ and fixed frequency $\omega=15$ and the variation is shown in Fig.~\ref{fig9}(b). For $\beta=1$ the value of $R$ is maximum however as $\beta $ decreases a drastic decrement in the area of the death region $R$ is seen. For $\beta<0.7$, area $R$ becomes zero, which implies restoration of oscillation in the entire parameter space. 

\begin{figure}[ht]
	\centerline{
		\includegraphics[scale=0.650]{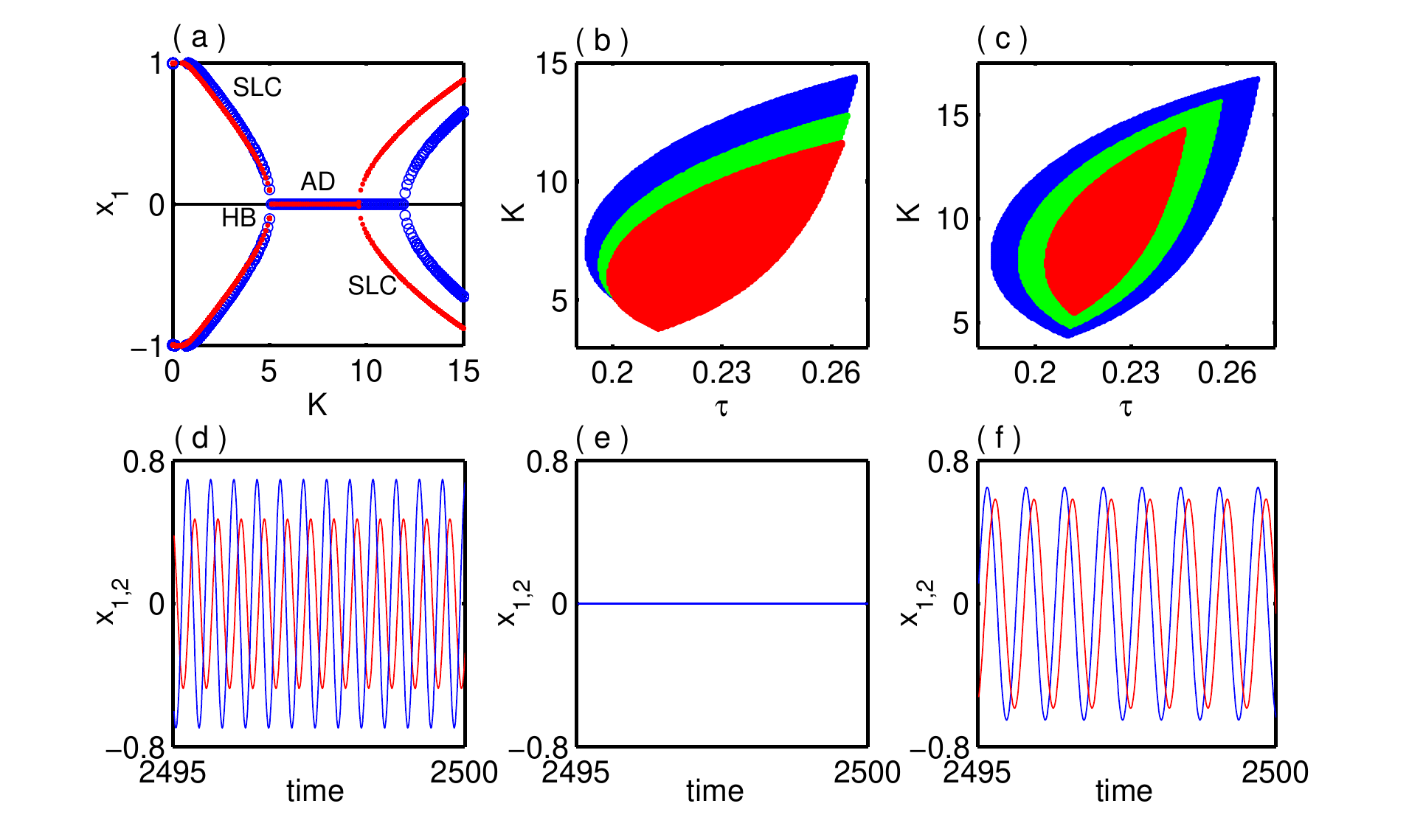}}
	\caption{(Color online) (a) The extrema of $x_1$ is plotted against the coupling strength $K$ for fixed $\beta=0.8$ and $\tau=0.2$ where the blue open circle and red dot are represents the corresponding values of $\alpha=0.4$ and $\alpha=0.5$.  (b) AD islands in $\tau-K$ parameter plane for $\beta=0.8$ and different values of $\alpha=0.5$ (blue), $\alpha=0.6$ (green), $\alpha=0.7$ (red). (c) The AD regions in $\tau-K$ parameter plane for $\beta=0.8$ (blue), $\beta=0.75$ (green), $\beta=0.7$ (red) keeping $\alpha=0.4$ fixed. The time series of (d) pre-death,  (e) death and (f) post death regimes for respective coupling strength $K=3.0, K=10.0$ and $K=16.0$ where $\tau=0.22,\alpha=0.4$ and $\beta=0.8$.}
	\label{fig10}
\end{figure}

\par  Next we testify the universality of our findings in the two coupled LS oscillators in the presence of both asymmetry and feedback parameters $\alpha$ and $\beta$ respectively. The results are shown in Fig.~\ref{fig10} consider different values of $\alpha$ and $\beta$. In Fig.~\ref{fig10}(a) the bifurcation diagram is plotted with respect to the coupling strength $K$ for fixed $\tau=0.2$, $\beta=0.8$ and two different values of $\alpha$ where the blue and red colors represent the extrema of $x_1$ at $\alpha=0.4$ and $\alpha=0.5$ respectively. The transition from stable limit cycle  to AD occurs at $K=5.2$ through HB and AD persists upto $K=11.8$ (blue open circle) for $\alpha=0.4$ and stable steady state is disappeared through inverse Hopf bifurcation. When $\alpha$ is slightly increased to $0.5$ the transition from SLC to AD occurs at the same value of $K$ but is sustained up to $K=9.5$ (red dot). Thus the regime of AD state at $\alpha=0.5$ is comparatively smaller than the AD state regime for $\alpha=0.4$ and in the remaining portions of the coupling strength oscillation is revived. The death islands are plotted in Fig.~\ref{fig10}(b) in the $\tau-K$ parameter space by taking different values of the asymmetry parameter $\alpha=0.5$ (blue), $\alpha=0.6$ (green) and $\alpha=0.7$ (red) and fixed values of feedback parameter $\beta=0.8$. From this figure it is clearly seen that the death region is shrinking in the $\tau-K$ space due to the increasing values of $\alpha$ with oscillations being restored in the remaining portion. Similarly, the variation of the death region is shown in Fig.~\ref{fig10}(c) for different values of $\beta$ and fixed value of $\alpha=0.4$. The blue, green and red colors represent the death area in the $\tau-K$ space for $\beta=0.8,\beta=0.75$ and $\beta=0.7$ respectively. In this case the death islands get contracted due to the decreasing values of the feedback parameter $\beta$ which is the opposite scenario observed in Fig.~\ref{fig10}(b). The death region becomes smaller because of increasing values of $\alpha$ when $\beta$ is fixed or due to decreasing values of $\beta$ when $\alpha$ is fixed where the corresponding oscillations are revived for the shrinking death region. From this discussion one can conclude that death and oscillatory regions can be monitored with proper tuning of the asymmetry parameter $\alpha$ and the feedback factor $\beta$. The time series $x_1,x_2$ of two coupled LS systems are plotted in Fig.~\ref{fig10} (d-f) while taking different values of coupling strength $K$ where the other parameters are fixed at $\tau=0.22,\alpha=0.4$ and $\beta=0.8$. If we draw a vertical line along $\tau=0.22$ in Fig.~\ref{fig10}(c) then at $K=3.0$ and $K=16.0$  there are no death islands while at $K=10.0$ death island is observed. The predeath and post death time series of $x_1,x_2$ are shown in Fig.~\ref{fig10}(d) and Fig.~\ref{fig10}(f) at $K=3.0$ and $K=16.0$ respectively. The time series at the AD state are shown in Fig.~\ref{fig10}(e) for $K=10.0$. From  Fig.~\ref{fig10}(d) and Fig.~\ref{fig10}(f), time series of the oscillations are limit cycle oscillations but their amplitude is different with some time lag. The amplitude variation happens due to the effect of both asymmetry and feedback parameter while the time lag is created due the presence of time delay in the interaction.
\begin{figure}[ht]
	\centerline{
		\includegraphics[scale=0.50]{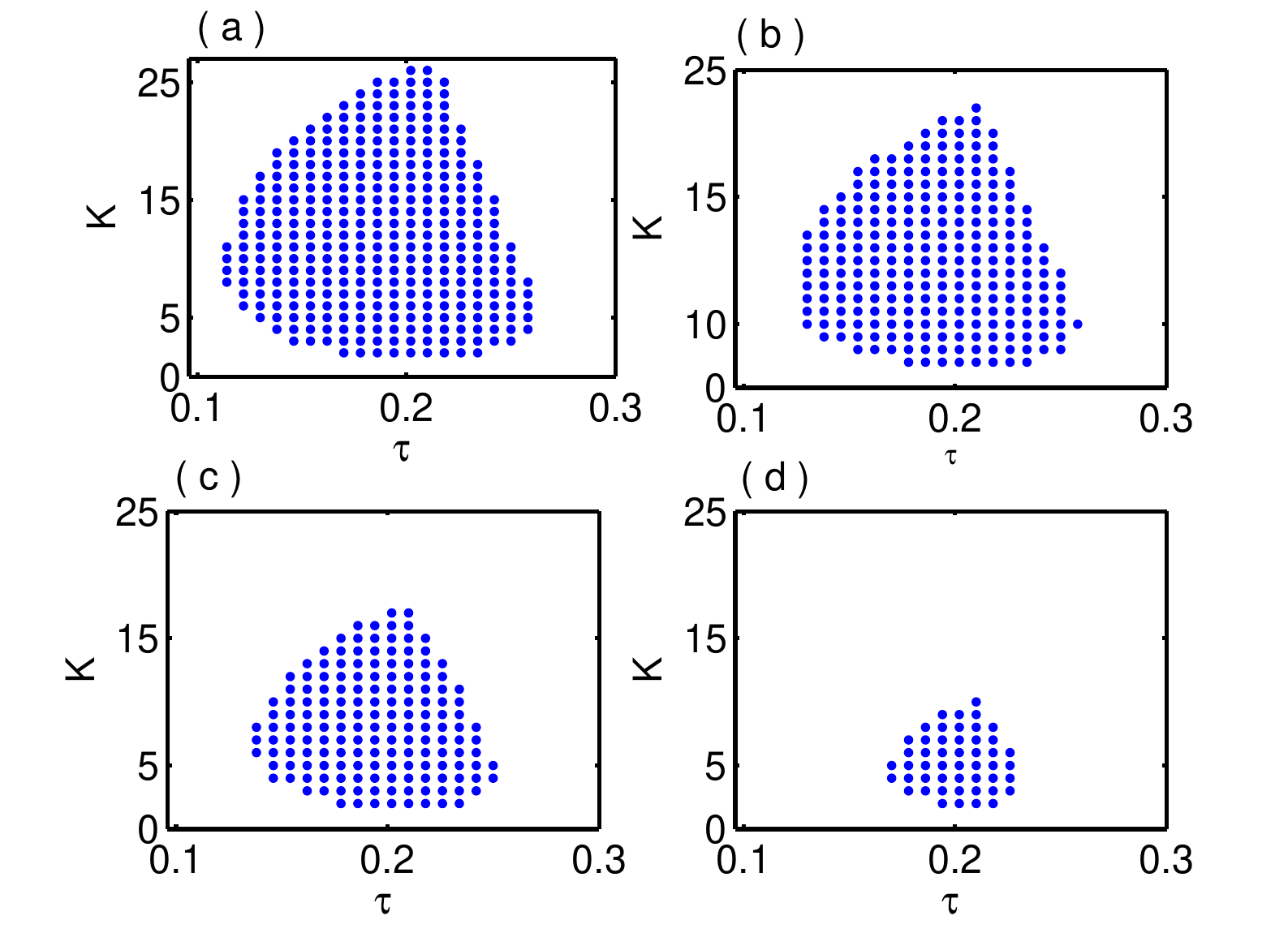}}
	\caption{ The amplitude death island of the network ($N=100$) of coupled oscillators in the space of coupling strength $k$ and time-delay $\tau$ for different values of feedback parameter $\beta$: (a) $\beta=0.95$, (b) $\beta=0.93$, (c) $\beta=0.9$ and (d) $\beta=0.8$. Here $\alpha=1.0.$ }
	\label{fig11}
\end{figure}
\par Now we consider network of oscillators in delayed cyclic form with feedback parameter $\beta$ involved in it to see how the area of the death region varies in such a network of coupled systems. In Fig.~\ref{fig11}, we take $N=100$ oscillators having the same frequency $\omega=15$ and different values of $\beta$ in $(K,\tau)$ parameter space. Fig.~\ref{fig11}(a-d) shows the AD regions with $ \beta=0.95, 0.93, 0.9$ and $0.8$ respectively. It is easily seen from Fig.~\ref{fig11}(a) that the AD appears for a broad range of parameters $K$ and $\tau$ due to high value of $\beta$ (namely, $\beta=0.95$). Decreasing $\beta$ value  to $0.93$ shrinks the death region for the parameter space shown in Fig.~\ref{fig11}(b). Further decrease in $\beta$ from $0.93$ to $0.9$ and $0.8$ decreases the  death region rapidly as seen in Fig.~\ref{fig11}(c) and Fig. ~\ref{fig11}(d) respectively, that precisely implies the resurrection of oscillation in a certain range of the $(K,\tau)$ parameter space. By continuing to decrease the value of $\beta$, for certain $\beta$ values the death region completely disappears and oscillation revives for any value of $K$ and $\tau$. This ensures that the death region decreases rapidly due to the decreasing values of the feedback $\beta$ even in case of network of many oscillators. 
\begin{figure}[ht]
	\centerline{
		\includegraphics[scale=0.50]{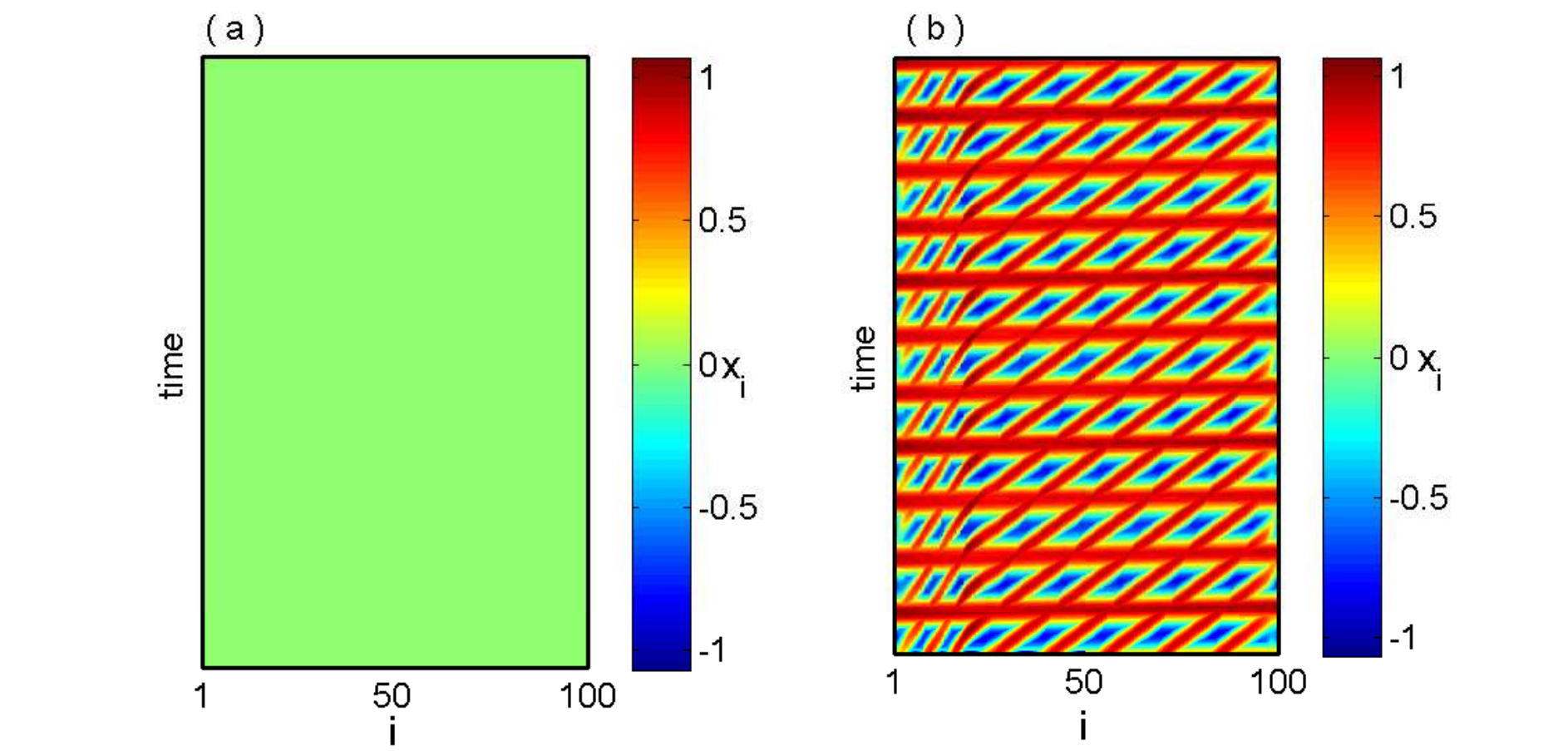}}
	\caption{(Color online) The spatio-temporal plots of amplitude death and oscillation revival regime in a network of coupled oscillators: (a) $\beta =1.0$ and (b) $\beta=0.6$. Here $K=10, \omega=15, \tau=0.15$ and $N=100$.}
	\label{fig12}
\end{figure}
The spatio-temporal plots corresponding to AD state and resurrected oscillatory states from AD are shown in Fig.~\ref{fig12}(a)  and Fig.~\ref{fig12}(b) for $\beta=1.0$ and   $\beta=0.6$ respectively. The Fig.~\ref{fig12}(a) shows that all of the $100$ oscillators of the network approach to the trivial fixed point (i.e., the origin) for the fixed values of $K=10, \omega=15, \tau=0.15$ at feedback parameter value $\beta=1$. By reducing the value of $\beta$ to $0.6$ all the oscillators in the network are oscillatory states from AD state and the spatio-temporal behavior is shown in Fig.~\ref{fig12}(b).
\begin{figure}[ht]
	\centerline{
		\includegraphics[scale=0.620]{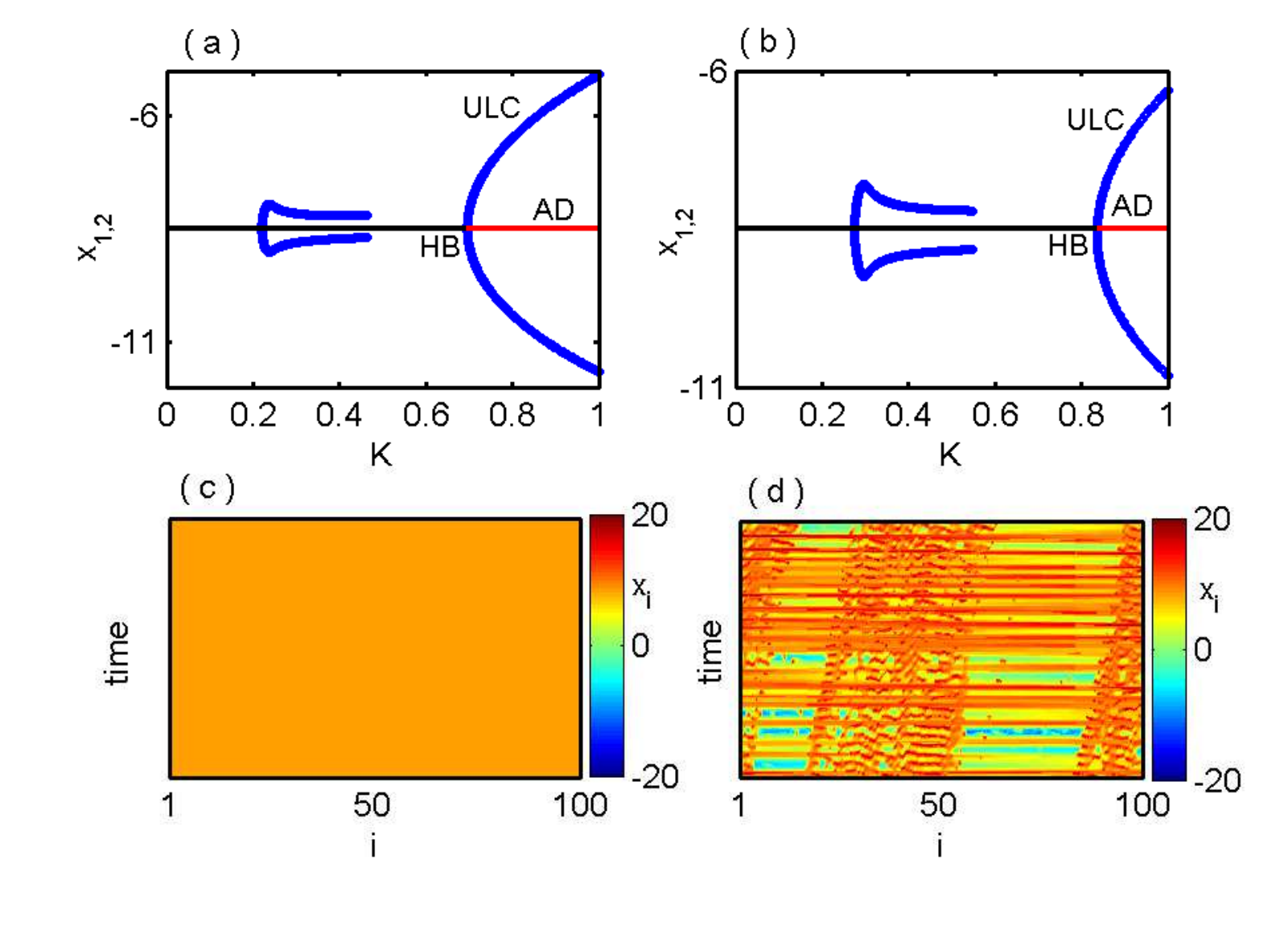}}
	\caption{(Color online) Two cyclic delay coupled Lorenz oscillators: Bifurcation diagrams with respect to $K$ for (a) $\beta=1$ and (b) $\beta=0.7$ at $\tau=1.0$, showing revival of oscillation from AD state by decreasing the value of $\beta$ from unity. The red solid line is for the stable fixed point, black line is for unstable fixed points and the blue open circle represents amplitude of unstable limit cycle that emerges through Hopf bifurcation. The spatiotemporal plots for $N=100$ coupled Lorenz oscillators: (c) AD state for $\beta=1$, (d) revival of desynchronized oscillation from AD state for $\beta=0.5$ with $K=5, \tau=1$. }
	\label{fig13}
\end{figure}

\section{ Coupled Lorenz oscillators }       
To confirm the revival of oscillation scenario from quenched states we extend our investigation into chaotic Lorenz systems. Here we will show that under delayed cyclic interaction, the oscillations are restored from the suppressed states by proper tuning of a feedback parameter in two coupled system as well as in the network of oscillators. The Lorenz oscillators in cyclic delayed form are described by the following equations:\\\\
$\dot x_i=a(y_i-x_i)+K(x_{i+1}-\beta x_i)\\\dot y_i=x_i(b-z_i)-y_i+K(y_{i-1}(t-\tau)-\beta y_i)\\\dot z_i=x_iy_i-cz_i$ $$ \eqno{(5)}$$
for $i=1,2,...,N (N\ge 3)$ with periodic boundary condition $ x_{N+1}(t)=x_1(t) $ and $ y_0(t-\tau)=y_N(t-\tau) $ as before in Fig.~\ref{fig1}. We choose the parameter values as $a=10$, $b=28$ and $c=\frac{8}{3}$ so that an individual Lorenz system exhibits chaotic behavior. Without coupling the individual oscillator oscillates chaotically and has a saddle point at the origin and another two unstable fixed points $(\pm\sqrt{c(b-1)},\pm\sqrt{c(b-1)},b-1)$. Here $\beta, \tau$ and $K$ are the feedback parameter, time delay and coupling strength respectively. It is obvious that the oscillations in coupled identical chaotic systems get suppressed if time delay is introduced in the interaction form. Here we elucidate how the feedback parameter revives oscillation from the quenched state. Bifurcation diagrams are plotted (using XPPAUTO \cite{xpp}) against the coupling strength $K$. In Fig.~\ref{fig13}(a), for sufficient time delay ($\tau=1$) in the interaction form, coupled systems produce AD for unit feedback ($\beta=1$) value. The AD emerges from unstable limit cycle (ULC) via Hopf bifurcation for critical values of coupling strength at $K=0.695$ but when we set the value of feedback parameter at $\beta=0.7$ then AD occurs at larger value of $K=0.835$ as seen in Fig.~\ref{fig13}(b) compared to the critical value of $K=0.695$ for $\beta=1$ in Fig.~\ref{fig13}(a). This effectively demonstrates the scenario of resurrection of oscillation from AD state.
\par Finally, we draw spatio-temporal plots referring to AD state and revived oscillatory state in Fig.~\ref{fig13}(c) and Fig. ~\ref{fig13}(d) with $\beta=1.0$ and $\beta=0.5$ respectively for a network of size $N=100$ where $K=5$ and $\tau=1$ are fixed.   

\section{Conclusion}
In this paper, we studied some collective behaviors, namely amplitude death and revival of oscillation from suppressed states of identically coupled systems under novel cyclic time delay interaction. Using asymmetry and feedback parameter, we can control the amplitude death island in a wide parameter space of coupling strength and time delay. By suitable characterization of death states and detailed bifurcation analysis the suppression and restoration processes of oscillations are analyzed. These two collective scenarios are observed in two coupled oscillators as well as in the network of large number of oscillators in limit cycle oscillators (Landau-Stuart) and chaotic oscillator (Lorenz). Described coupling schemes are usually seen in neuronal systems and such type of interaction may also occur in many physical, biological and engineering systems.  \\\\ 
\\
{\bf Author contribution statement}\\
B.K.B. and S.M. have carried out the numerical and analytical calculations. All three authors have equally contributed to the discussion, the interpretation of the results and the writing of the manuscript.

\end{document}